\documentclass[a4paper,10pt,twoside]{cpc-hepnp}

\usepackage{multicol}
\usepackage{graphicx}
\usepackage{booktabs}
\usepackage{amssymb,bm,mathrsfs,bbm,amscd}
\usepackage[tbtags]{amsmath}
\usepackage{lastpage}

\begin{document}

\fancyhead[co]{\footnotesize Huang Ming-Yang: Study on accelerator neutrino detection at a spallation source}


\title{Study on accelerator neutrino detection at a spallation source\thanks{Supported by National Natural
Science Foundation of China (Nos.11205185 and 11175020)}}

\author{%
      Huang Ming-Yang$^{1,2;1)}$\email{huangmy@ihep.ac.cn}%
}
\maketitle

\address{%
$^1$ Institute of High Energy Physics (IHEP), Chinese Academy of Sciences (CAS),
Beijing 100049, China\\
$^2$ Dongguan Institute of Neutron Science (DINS), Dongguan 523808, China\\
}

\begin{abstract}

In this paper, we study the detection of accelerator neutrinos produced at the
China Spallation Neutron Source (CSNS). Using the code FLUKA, we have simulated
the production of neutrinos in the proton beam on the tungsten target and obtained
the yield efficiency, numerical flux, and average energy of different flavors
of neutrinos. Furthermore, detections of these accelerator neutrinos are investigated
in two reaction channels, the neutrino-electron reactions and neutrino-carbon
reactions. The event numbers of different flavors of neutrinos have also been
calculated.

\end{abstract}




\begin{multicols}{2}

\section{Introduction}

In the past several decades, a number of spallation neutron sources have started
to operate, such as the Los Alamos Meson Physics Facility (LAMPF),
the Spallation Neutron Source at Rutherford Appleton Laboratory (ISIS) \cite{ISIS1},
the Japan Accelerator Research Complex (J-PARC) \cite{J-PARC1},
the Spallation Neutron Source at Oak Ridge National Laboratory (SNS) \cite{SNS1},
etc.  In recent years, new spallation neutron sources are under construction, such
as the China Spallation Neutron Source (CSNS) \cite{CSNS2}, the European Spallation
Neutron Source (ESS) \cite{ESS1}, etc. These spallation neutron sources are
designed to provide wide multidisciplinary platforms for scientific
research and industrial applications at national institutes, universities,
and industrial laboratories \cite{Wei1, Wei2}. The areas concerned include basic energy
sciences, particle physics, and nuclear sciences. Table 1 shows the main technical
parameters of several major spallation neutron sources at the GeV energy range. For
J-PARC, the technical parameters shown in the table refer to the Rapid Cycling
Synchrotron (RCS).

\end{multicols}

\begin{center}
\tabcaption{ \label{tab1} Main technical parameters for several spallation neutron sources
at the GeV energy range (¡°ppp¡± means protons per pulse).}
\footnotesize
\begin{tabular*}{180mm}{@{\extracolsep{\fill}}l|cccccc}
\toprule    & Extraction energy & Extraction power & Repetition rate & Average beam Current & Intensity & Target  \\
           & (GeV) & (MW) & (Hz) & (mA) & ($10^{13}$ppp) &   \\
 \hline
 LAMPF \cite{LSND1}   & 0.8   & 0.056    &  120 & 1     & 2.3       &  Various \\
 ISIS  \cite{ISIS2, ISIS3}    & 0.8   & 0.16     & 50   & 0.2   & 2.5        &  Water cooled       \\
         &       &         &       &      &            &   /Tantalum      \\
 J-PARC \cite{Shirakata1, Saha1}   & 3.0   & 1.0      & 25   & 0.333 & 8.3        &  Mercury      \\
 SNS \cite{SNS2, VanDalen1}  & 1.0   & 1.4      & 60   & 1.6   & 16       &  Mercury      \\
 CSNS-I(II) \cite{Wang1, Huang2}& 1.6   & 0.1(0.5) & 25   & 0.063(0.315)     & 1.56(7.8)  &  Tungsten       \\
 ESS  \cite{ESS2}     & 2.0   & 5.0      & 14   & 62.5  & 110       &  Tungsten      \\
\bottomrule
\end{tabular*}%
\end{center}

\begin{multicols}{2}

The CSNS consists of an 80 MeV proton linac, a 1.6 GeV RCS, a solid tungsten target
station, and various instruments for applications spallation neutrons \cite{Wei3}.
The accelerator operates at 25 Hz repetition rate with
an initial design beam power of 100 kW and can be upgraded to 500 kW.
As the exclusive spallation neutron source in developing countries, CSNS will be
among the top four of such facilities in the world upon completion. Table 2 shows
the main design parameters of CSNS-I and CSNS-II \cite{Wang1,Huang2}.

A large number of neutrinos can be produced at the beam stops of high intensity
proton accelerators.  They form the neutrino beams for basic scientific studies
to better understand properties of neutrinos and better probes of
the weak interaction force \cite{Kopp1}. During the last several decades, many
neutrino experiments have performed based on neutrino beams from spallation
neutron sources \cite{Burman1} and other proton accelerators.  They include the
Liquid Scintillator Neutrino
Detector (LSND) \cite{LSND1} at LAMPF, the Karlsruhe Rutherford Medium Energy
Neutrino experiment (KARMEN) \cite{KARMEN1, KARMEN4} at ISIS,
the Tokai-to-Kamioka experiment (T2K) \cite{T2k1} at J-PARC and so on. For CSNS,
similar to other accelerators, an intensive beam of accelerator neutrinos can be
brought by the proton beam hitting on the tungsten target. By using a neutrino
detector similar to MiniBooNE \cite{MiniBooNE1}, different flavors of neutrinos
can be detected. Therefore, neutrino properties, such as neutrino mixing parameters
and the mass hierarchy may be studied.

\begin{center}
\tabcaption{ \label{tab2} Main design parameters of CSNS.}
\footnotesize
\begin{tabular*}{80mm}{@{\extracolsep{\fill}}l|cc}
\toprule Parameter/unit    & CSNS-I & CSNS-II  \\
 \hline
 Beam power on target/MW & 0.1   &  0.5   \\
 Linac energy/GeV  &    0.08    &   0.25    \\
 Beam energy on target/GeV & 1.6 & 1.6   \\
 Average beam current/$\mu$A  & 62.5  &  315  \\
 Pulse repetition rate/Hz  &  25  &  25      \\
 Ion type, source linac  &  $H^-$  &  $H^-$        \\
 Protons per pulse/$10^{13}$    &  1.56    &   7.8   \\
 Target material   &  Tungsten  &  Tungsten   \\
 Target number   &  1  &  1   \\
 Target size/mm$^3$ & $50\times150\times400$   &  $50\times150\times400$   \\
 Beam section/mm$^2$  &    $40\times100$    &  $40\times100$    \\
\bottomrule
\end{tabular*}%
\end{center}

\section{Accelerator Neutrino Beam}

The idea of using accelerator neutrino beam to study neutrinos was initiated
independently by Schwartz and
Pontecorvo, and the experiment was first carried out by Lederman, Schwartz,
Steinberger and collaborators \cite{Kopp1, Bonesini1}. Low energy neutrino beams
can be produced by the decays of $\pi$ and $\mu$ at rest, which are generated in
low energy proton beams hitting on targets. In a beam dump experiment,
the target for the primary proton beam, where the neutrino parent particles
emerge, is the medium for absorbing or stopping the hadrons. Then
no drift space is provided for hadrons to decay in. Therefore, high intensity and
low energy proton accelerators ($p_{beam}\sim1$ GeV/c) such as spallation neutron
sources are commonly used \cite{Bulanov1, Lazauskas1}.
The two recent experiments at a beam dump, LSND and KARMEN, have given
controversial results on neutrino oscillations which have been resolved
by the MiniBooNE experiment \cite{Louis1, MiniBooNE2}.

The dominant decay scheme that produces neutrinos from a stopped pion source is \cite{Gervey1}
\begin{equation}
 \pi^+\longrightarrow\mu^++\nu_{\mu}, \quad \tau_{\pi}=26 ns,
 \label{pidecay}
\end{equation}
followed by
\begin{equation}
 \mu^+\longrightarrow e^++\overline{\nu}_{\mu}+\nu_e, \quad \tau_{\mu}=2.2 \mu s,
 \label{nudecay}
\end{equation}
where $\tau_{\pi}$($\tau_{\mu}$) is the life time of $\pi^+$($\mu^+$).

The bulk of the $\pi^-$'s generated are strong absorbed by the target before they
are able to decay,
and most of the $\mu^-$'s produced from the $\pi^-$ decay are captured from
the atomic orbit, a process which does not give rise to $\bar{\nu}_e$.
Therefore, the yield efficiency of $\bar{\nu}_e$ is a factor of $10^{-3}$ to
$10^{-4}$ lower, and it will be neglected in the following discussions.

\begin{center}
\begin{tabular}{ccccc}
\scalebox{0.45}{\includegraphics{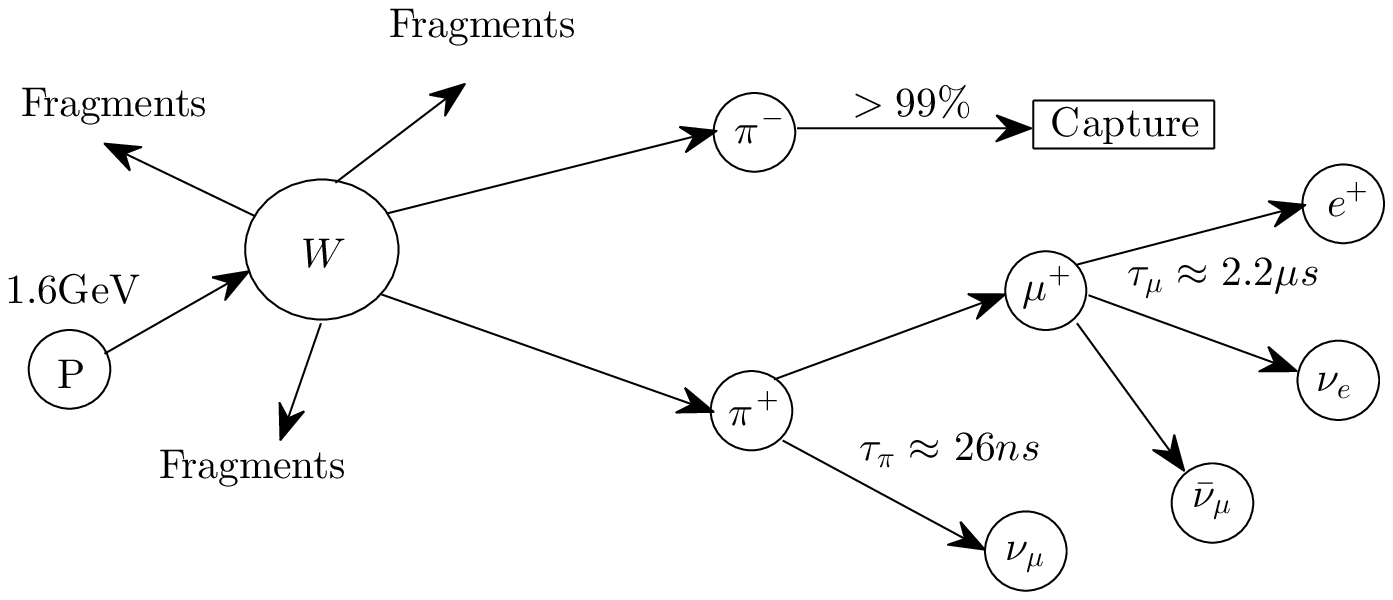}}\\
\end{tabular}\\
\figcaption{\label{production}
The production mechanism of accelerator neutrinos at CSNS.}
\end{center}

The CSNS beam stop will provide a copious flux of neutrinos, primarily
from $\pi^+$ and $\mu^+$ decays. Fig. 1 shows the scheme of neutrino production
in the tungsten beam stop. By using the code FLUKA \cite{Ferrari1} and the main design parameters
given in Table 2, the processes that the 1.6 GeV proton beam hitting
the tungsten target were simulated. The results show that, after the complete decays of
$\pi^+$ and $\mu^+$, three different species of neutrinos ($\nu_e$,
$\nu_{\mu}$, $\bar{\nu}_{\mu}$) produced have the same yield efficiency which
is about 0.17 per proton.
In Table 3, the event numbers per year of accelerator neutrinos produced
at CSNS are given. We find that the three species of accelerator neutrinos have
the same yield event number per year which is $0.21\times10^{22}$ for CSNS-I and $1.05\times10^{22}$ for CSNS-II.
Since it will rise ultimately to 500 kW in the future with
more neutrinos generated, the main parameters for CSNS-II
will be considered mainly in the following research.

\begin{center}
\tabcaption{ \label{tab3} Event numbers per year of accelerator neutrinos
produced at CSNS.}
\footnotesize
\begin{tabular*}{80mm}{@{\extracolsep{\fill}}l|cc}
\toprule        & CSNS-I & CSNS-II  \\
 \hline
 Proton number per year/$10^{22}$ & 1.23 & 6.15   \\
 $\nu_e$ number per year/$10^{22}$ & 0.21  &  1.05  \\
 $\nu_{\mu}$ number per year/$10^{22}$ &  0.21  &  1.05   \\
 $\bar{\nu}_{\mu}$ number per year/$10^{22}$ &  0.21  &  1.05        \\
\bottomrule
\end{tabular*}%
\end{center}

The numerical flux $\phi$ of each particle ($\pi^+$, $\mu^+$, $\nu_e$,
$\nu_{\mu}$, $\bar{\nu}_{\mu}$) can be found in \cite{Vergados1}

\begin{equation}
 \phi(L) = \frac{\varphi(num/year)}{4\pi L^2 (cm^2)},
 \label{Flux}
\end{equation}
where $L$ is the distance of the spallation target from the neutrino detector,
and $\varphi$ is the yield event number per year. By using FLUKA, the yield
efficiency of different particles can be obtained, then the numerical flux can
be calculated using Eq. (\ref{Flux}).
Fig. 2 shows that:

(i) Numerical fluxes of various particles involved decrease with the distance
$L$;

(ii) $\nu_e$ and $\bar{\nu}_{\mu}$ have the same numerical flux;

(iii) The numerical flux of $\nu_{\mu}$
is larger than those of $\nu_e$ and $\bar{\nu}_{\mu}$;

(iv) The numerical flux of $\pi^+$ decreases very quickly with the distance $L$.

\begin{center}
\begin{tabular}{ccccc}
\scalebox{0.38}{\includegraphics{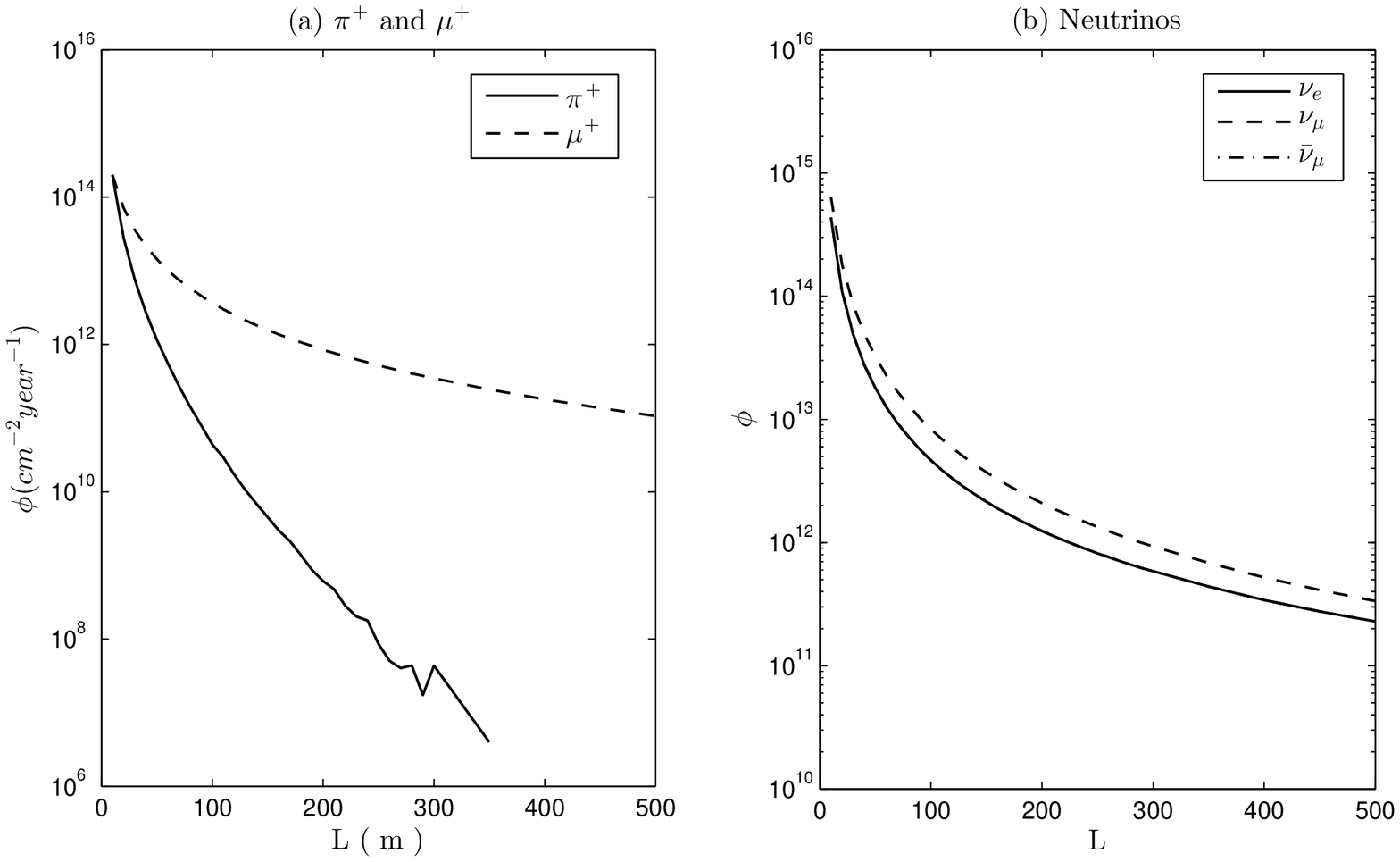}}\\
\end{tabular}\\
\figcaption{\label{flux}
The numerical flux of each particle ($\pi^+$, $\mu^+$, $\nu_e$,
$\nu_{\mu}$, $\bar{\nu}_{\mu}$) as a function of the distance $L$.
(a) $\pi^+$ and $\mu^+$; (b) $\nu_e$,
$\nu_{\mu}$, and $\bar{\nu}_{\mu}$.}
\end{center}

From the simulation results of neutrino productions, the average energy of
different species of neutrinos can be obtained. The energy spectra of the
different speies of neutrinos are given in Fig. 3. It can be obtained that the average
energy of $\nu_e$ is much smaller than that of $\nu_{\mu}$ and $\bar{\nu}_{\mu}$.

\begin{center}
\begin{tabular}{ccccc}
\scalebox{0.38}{\includegraphics{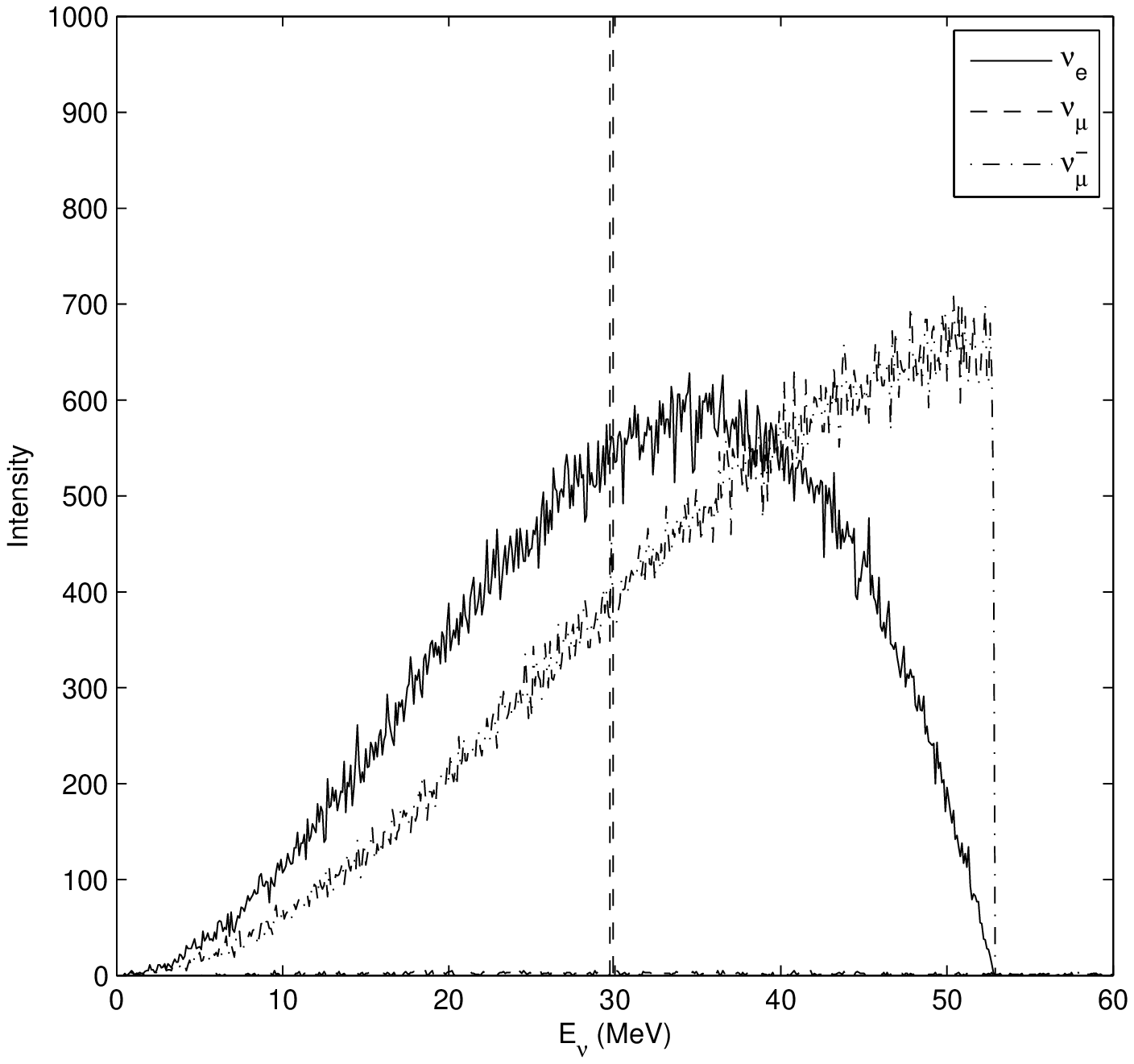}}\\
\end{tabular}\\
\figcaption{\label{spectra}
Accelerator neutrino energy spectra at CSNS.}
\end{center}

Because of the flavor mixing there are oscillations among $\nu_e$, $\nu_\mu$,
and $\nu_{\tau}$ \cite{Olive1, Huang4, Huang5}. In the three flavor mixing
scheme, neglecting the matter effect, and using the fact
$\Delta m^2_{21}=7.5\times10^{-5}$ eV$^2$ and $L/E_{\nu}\backsimeq1$, hence
$\sin^2(\Delta m^2_{21}\cdot L/4E_{\nu}) \backsimeq 0 $, we can write the
oscillation probabilities \cite{Dore1}
\begin{eqnarray}
 P(\nu_{e}\rightarrow\nu_{\mu}) \simeq \sin^2\theta_{23} \sin^22\theta_{13}
 \sin^2(\Delta m^2_{32}\cdot L/4E_{\nu}), \nonumber\\
 P(\nu_{e}\rightarrow\nu_{\tau}) \simeq \cos^2\theta_{23} \sin^22\theta_{13}
 \sin^2(\Delta m^2_{32}\cdot L/4E_{\nu}), \label{P}\\
 P(\nu_{\mu}\rightarrow\nu_{\tau}) \simeq \cos^4\theta_{13} \sin^22\theta_{23}
 \sin^2(\Delta m^2_{32}\cdot L/4E_{\nu}),
 \nonumber
\end{eqnarray}
where $\mid\Delta m^2_{32}\mid=2.4\times10^{-3}$ eV$^2$, $\sin^2\theta_{23}=0.446$,  $\sin^2\theta_{13}=0.0237$ \cite{Gapozzi1},
$E_{\nu}$ is the neutrino energy, and $L$ is the distance of the neutrino
source from the detector. Given the distance $L=60$ m the simulation results show
that the average energy of $\nu_e$, $\nu_{\mu}$, and $\bar{\nu}_{\mu}$ are 33.0 MeV,
50.0 MeV, 47.9 MeV respectively.
The oscillation probabilities can be calculated as
$P(\nu_{e}\rightarrow\nu_{\mu})\backsimeq1.27\times10^{-6}$, $P(\nu_{e}\rightarrow\nu_{\tau})\backsimeq1.57\times10^{-6}$,
and $P(\nu_{\mu}\rightarrow\nu_{\tau})\backsimeq1.26\times10^{-5}$. Therefore,
due to the very short baseline, the oscillations among $\nu_e$,
$\nu_\mu$, and $\nu_{\tau}$ can be neglected and will not enter our discussions
below.

In the next section, processes of the accelerator neutrino detection will be
studied, and the corresponding neutrino event numbers observed through various
reaction channels will be calculated.

\section{Detection of accelerator neutrinos}

The event numbers per year $\tilde{N}_i$ of accelerator neutrinos observed through
various reaction channels ¡°$i$¡± can be calculated following \cite{Elnimr1}

\begin{equation}
 \tilde{N}_i = \phi(\nu/year/cm^2)\cdot \sigma_i(cm^2)\cdot N_{T},
 \label{Event}
\end{equation}
where $\phi$ is the neutrino numerical flux given in Eq. (\ref{Flux}),
$\sigma_i$ is the cross section of the
given neutrino reaction, and $N_{T}$ is the target number.

For the accelerator neutrino detection at CSNS, a detector similar to MiniBooNE \cite{MiniBooNE1} is adopted. It consists of a spherical 803 tons fiducial mass
of mineral oil (CH$_2$, density 0.845 g/cm$^3$) and has a fiducial
radius of 6.1 m, occupying a volume of 950 m$^3$.
The total numbers of target protons, electrons, and $^{12}C$ are
\begin{eqnarray}
&&\ N_T^{(p)}=6.90\times10^{31},
\quad N_T^{(e)}=2.76\times10^{32}, \nonumber \\
&&\ N_T^{(C)}=3.45\times10^{31}. \nonumber
\end{eqnarray}

According to the discussions given in the previous section, the yield efficiency
of $\bar{\nu}_e$ is a factor of $10^{-3}$ to $10^{-4}$ lower than that of the
other neutrino species, and hence can be neglected. Then, the reaction channel
of the inverse beta decay will be neglected. For the neutrino-proton elastic
scattering, due to the Cerenkov energy threshold and quenching, only the high
energy part of the neutrino spectra can be observed and it has always been taken
to be too small in number to observe. In addition, due to
the the proton structure, protons in the neutrino-proton elastic scattering are
too difficult to identify and have large systematic uncertainty \cite{Beacom2}. Therefore, the neutrino-proton
elastic scattering will be not considered in this paper.

By using the said detector at CSNS, two reaction channels will be used to detect
the different species of neutrinos ($\nu_e$, $\nu_{\mu}$, and $\bar{\nu}_{\mu}$):

\vspace{3mm}

(1) Neutrino-electron reactions
\begin{eqnarray}
&&\ \nu_{e}+e^-\rightarrow \nu_{e}+e^- \quad({\rm CC\quad and\quad NC}), \nonumber\\
&&\ \nu_{\mu}+e^-\rightarrow \nu_{\mu}+e^- \quad({\rm NC}), \nonumber\\
&&\ \bar{\nu}_{\mu}+e^-\rightarrow \bar{\nu}_{\mu}+e^-
\quad({\rm NC}), \nonumber
\end{eqnarray}
where CC and NC stand, respectively, for the changed-current and
neutral-current interactions, producing recoil electrons with energy
from zero up to the kinematics maximum. The neutrino events observed through
these reaction channels can be identified by the signal of the recoil electrons
which are strong peaked along the neutrino direction \cite{Allen2, Imlay1},
and this forward peaking is usually used for experiments to distinguish the
electron elastic scattering from the neutrino reactions on nuclei.

\vspace{3mm}

(2) Neutrino-carbon reactions

\vspace{3mm}

For the neutrinos and $^{12}C$ system, there are one
charged-current and three neutral-current reactions:

Charged-current capture of $\nu_e$:
\begin{eqnarray}
&&\ \nu_e+^{12}C\rightarrow^{12}N+e^-, \quad E_{th}=17.34MeV,
\nonumber
\\
&&\ ^{12}N\rightarrow^{12}C+e^{+}+\nu_e.
\nonumber
\end{eqnarray}

Neutral-current inelastic scattering of $\nu_e$, $\nu_{\mu}$, and $\bar{\nu}_{\mu}$:
\begin{eqnarray}
&&\ \nu_{e}+^{12}C\rightarrow^{12}C^{\ast}+\nu_{e}^{'},
\quad E_{th}=15.11MeV,
\nonumber \\
&&\ \nu_{\mu}+^{12}C\rightarrow^{12}C^{\ast}+\nu_{\mu}^{'},
\quad E_{th}=15.11MeV,
\nonumber \\
&&\
\bar{\nu}_{\mu}+^{12}C\rightarrow^{12}C^{\ast}+\bar{\nu}_{\mu}^{'},
\quad E_{th}=15.11MeV, \nonumber\\
 &&\ ^{12}C^{\ast}\rightarrow^{12}C+\gamma.
\nonumber
 \end{eqnarray}
The charged-current events have the delayed coincidence of a $\beta$ decay following
the interaction.
The neutral-current events have a monoenergetic $\gamma$ ray at $15.11$ MeV. Therefore,
the charged-current and neutral-current reactions on carbon can be identified and
observed by the neutrino detector \cite{Auerbach1, Cadonati1}.

The effective cross sections of the above two reactions, the neutrino-electron
reactions \cite{Cadonati1, Arafune1} and
neutrino-carbon reactions \cite{Fukugita1, Kolbe5}, are given in Fig. 4.

The neutrino event numbers per year can be calculated using Eqs. (\ref{Flux})
and (\ref{Event}). Fig. 5 shows the neutrino event numbers per year
observed through two reaction channels, changing with the distance $L$.
We found that:

(i) The event numbers per year of different species of neutrinos all decrease
with the distance $L$ for both the neutrino-electron reactions and neutrino-carbon reactions;

(ii) The total event number of accelerator neutrinos observed through the
neutrino-carbon channel is much larger than through the neutrino-electron
reactions;

(iii) For the neutrino-electron reactions,
$\tilde{N}_{\nu_e}(CC+NC)>\tilde{N}_{\nu_{\mu}}(NC)>\tilde{N}_{\bar{\nu}_{\mu}}(NC)$;

(iv) For the neutrino-carbon reactions, $\tilde{N}_{\nu_{\mu}}(NC)>\tilde{N}_{\bar{\nu}_{\mu}}(NC)>\tilde{N}_{\nu_e}(NC)$; $\tilde{N}_{\nu_{\mu}}(NC)$ and $\tilde{N}_{\nu_e}(CC+NC)$ are both
larger than $\tilde{N}_{\bar{\nu}_{\mu}}(NC)$.

\begin{center}
\begin{tabular}{ccccc}
\scalebox{0.34}{\includegraphics{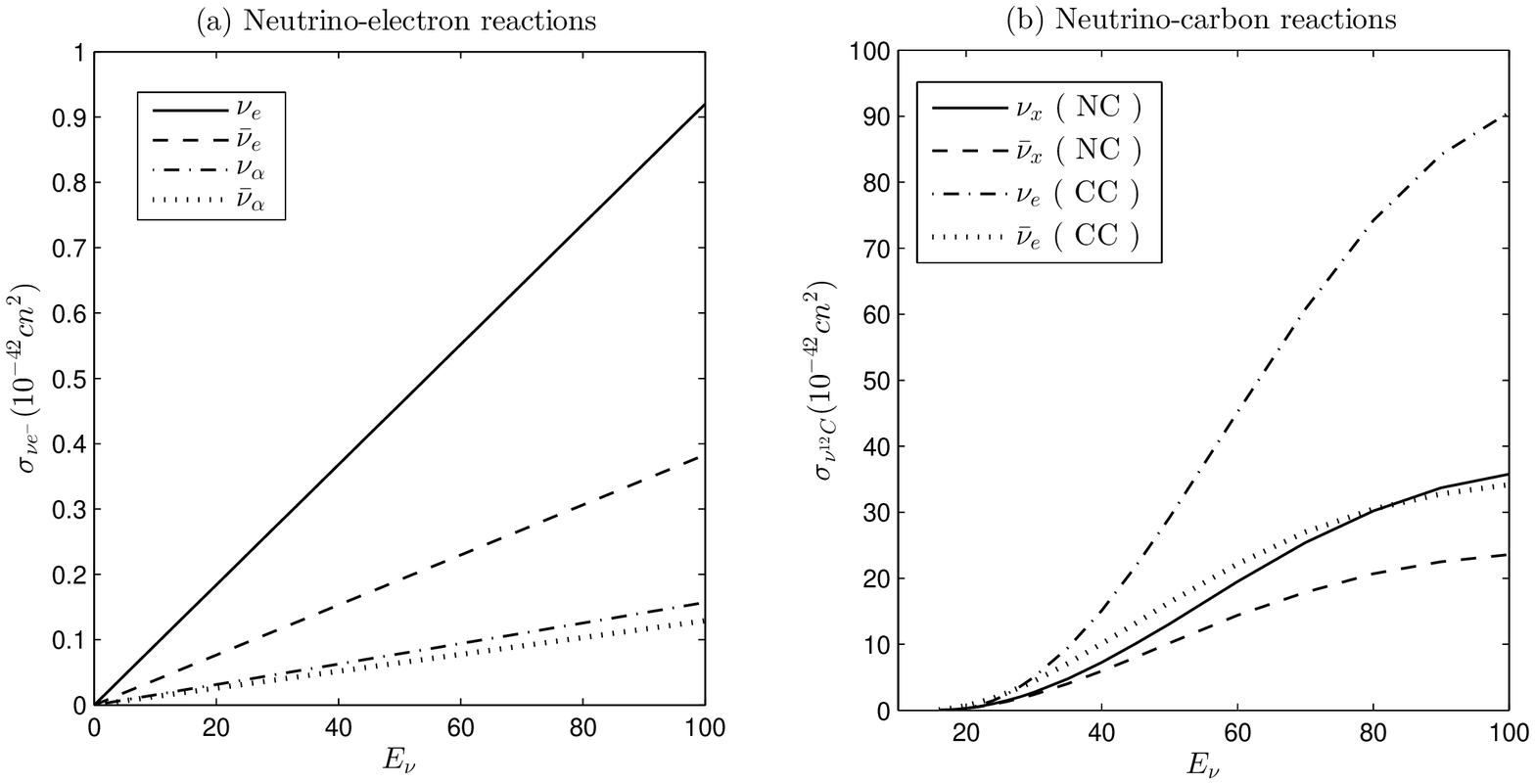}}\\
\end{tabular}\\
\figcaption{\label{CS}
 The effective cross sections as functions of the neutrino energy.
(a) the neutrino-electron reactions; (b) the neutrino-carbon reactions.
$\alpha=\mu, \tau$, $x=e, \mu, \tau$.}
\end{center}

\begin{center}
\begin{tabular}{ccccc}
\scalebox{0.4}{\includegraphics{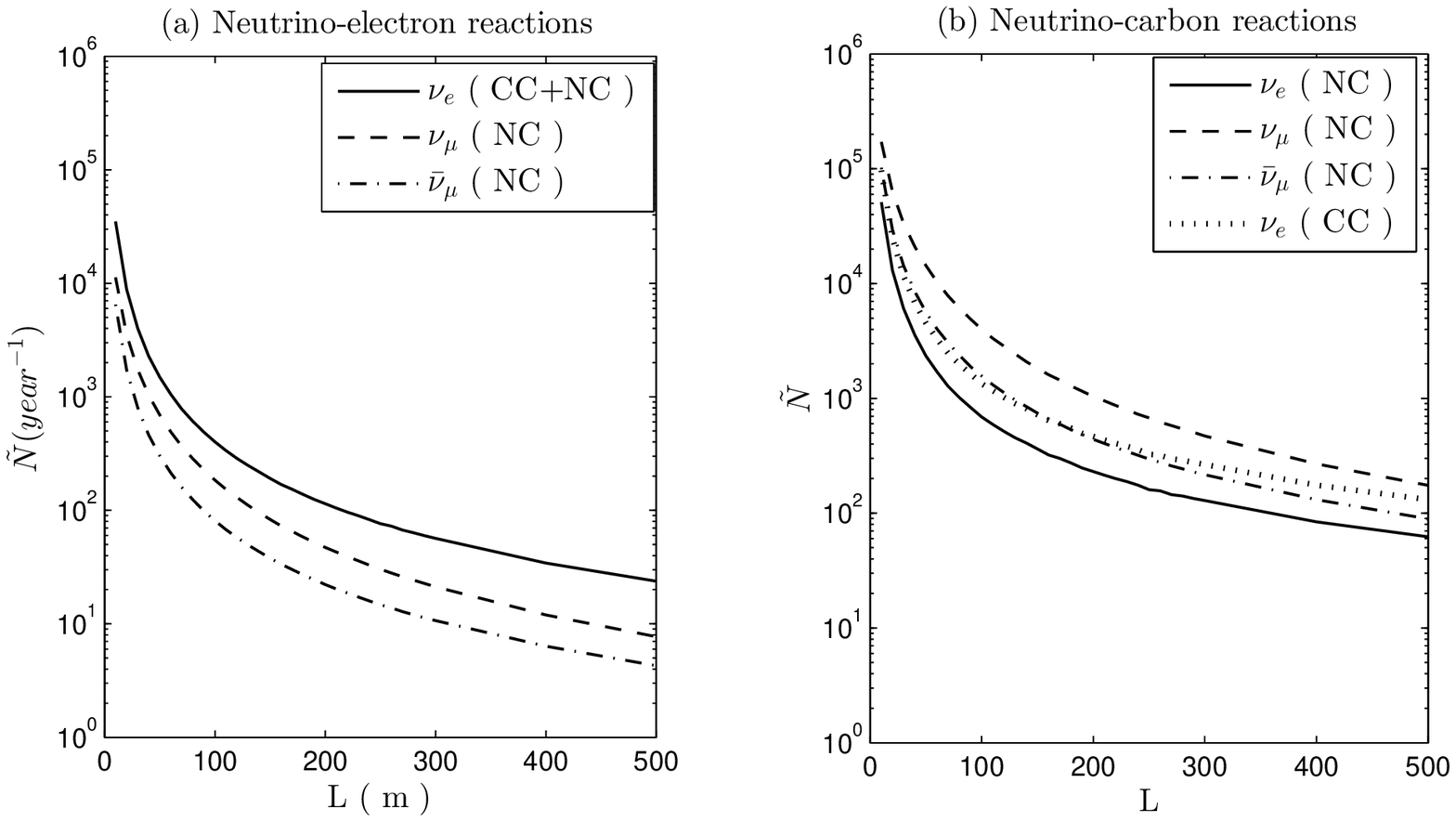}}\\
\end{tabular}\\
\figcaption{\label{number}
 The event numbers per year of accelerator neutrinos
observed through various reaction channels as functions of the distance $L$.
(a) the neutrino-electron reactions; (b) the neutrino-carbon reactions.}
\end{center}

\begin{center}
\tabcaption{ \label{tab4} Neutrino event numbers per year observed through various reaction channels for $L=60$ m.  The detector efficiency and beam-on efficiency are
both taken to be 50\%.}
\footnotesize
\begin{tabular*}{80mm}{@{\extracolsep{\fill}}c|cccc}
\toprule Reaction & $\tilde{N}_{\nu_e}(CC)$ $+$ $\tilde{N}_{\nu_e}(NC)$ & $\tilde{N}_{\nu_{\mu}}(NC)$ & $\tilde{N}_{\bar{\nu}_{\mu}}(NC)$  \\
 \hline
 $\nu e^-$    &       263       &  123    &    54     \\
 $\nu^{12}C$  &   824  +  426   &  2580   &    1005    \\
\bottomrule
\end{tabular*}%
\end{center}

In the future, if the distance $L$, the detector efficiency, and the beam-on
efficiency are defined, the neutrino event numbers per year observed through
various reaction channels can be calculated accurately.
For example, suppose the distance $L=60$ m, the detector efficiency and beam-on
efficiency both at $50\%$, the accurate neutrino event numbers per year
are given in Table 4. It is clear that
there are a large number of accelerator neutrinos
which can be used for measuring neutrino cross sections.

\section{Summary and Discussion}

In this paper, the accelerator neutrino beam has been studied in detail.
With the code FLUKA, processes of accelerator neutrino production at CSNS
from the proton beam on the tungsten target have been investigated,
and the yield efficiency, numerical flux, average energy of different species
of neutrinos have been obtained. We show that, after the complete decays of
$\pi^+$ and $\mu^+$, three kinds of accelerator neutrinos have the same yield
efficiency which is about 0.17 per proton. Therefore, they have the same yield
event number per year which is $0.21\times10^{22}$ for CSNS-I and
$1.05\times10^{22}$ for CSNS-II.

The detection of accelerator neutrinos through two reaction channels,
the neutrino-electron reactions and neutrino-carbon reactions, has been
studied, and the neutrino event numbers have been calculated. It is found that
the total event number of accelerator neutrinos observed through the
channel of the neutrino-carbon reactions is much larger than that of the
neutrino-electron reactions.

In our calculation, the detector efficiency and beam-on efficiency were
not seriously considered. In the future, in the completion of the design
of a neutrino detector at CSNS, the detector efficiency
and beam-on efficiency will be given. Then, the detector errors will need
to be reduced in the calculation of the neutrino event numbers. Furthermore,
the statistical errors and systematic errors on neutrino fluxes and cross
sections also need to be considered for the design of neutrino experiment.

\vspace{10mm}

\acknowledgments{The author would like to thank B.-L. Young, X.-H. Guo, S. Wang,
and S.-J. Ding for helpful discussions and support.}

\end{multicols}

\vspace{10mm}

\vspace{-1mm}
\centerline{\rule{80mm}{0.1pt}}
\vspace{2mm}

\begin{multicols}{2}

\end{multicols}

\clearpage

\end{document}